\documentclass{aa}
\usepackage[varg]{txfonts}
\usepackage{amsmath}
\usepackage{graphicx}
\usepackage{url}
\usepackage{listings}
\usepackage{xcolor}
\usepackage{tabu}
\usepackage{units}
\usepackage{wasysym}
\usepackage[font={small},labelfont=bf,belowskip=2mm,aboveskip=1mm]{caption}
\usepackage{geometry}
\usepackage{multirow}
\usepackage{leftidx}
\usepackage[hidelinks]{hyperref}
\usepackage{csquotes}
\bibpunct{(}{)}{;}{a}{}{,} 

\def\teff{T_\text{eff}}

\def\logg{\log{g}}
\def\feh{\left[ \mathrm{Fe} / \mathrm{H} \right]}
\def\mh{\left[ \mathrm{M} / \mathrm{H} \right]}
\def\afe{\left[ \alpha / \mathrm{Fe} \right]}

\def\snr{\mathrm{S} / \mathrm{N}}

\title{Estimating $\afe$ from Gaia low-resolution BP/RP spectra using the ExtraTrees algorithm}
\titlerunning{Estimating $\afe$ from BP/RP spectra}
\author{Alvin~Gavel\inst{\ref{inst:uppsala}}, Ren{\'e} Andrae\inst{\ref{inst:mpia}}, Morgan Fouesneau\inst{\ref{inst:mpia}}, Andreas J. Korn\inst{\ref{inst:uppsala}}, Rosanna Sordo\inst{\ref{inst:padova}}}
\institute{Observational Astrophysics, Division of Astronomy and Space Physics, Department of Physics and Astronomy, Uppsala University, Box 516, 75120 Uppsala, Sweden\label{inst:uppsala}
\and
Max-Planck-Institute for Astronomy, K\"onigstuhl 17, 69117 Heidelberg, Germany\label{inst:mpia}
\and
INAF-Osservatorio Astronomico di Padova, Vicolo Osservatorio 5, 35122, Padova, Italy\label{inst:padova}
}
\date{Received 18 June, 2021}

\abstract
{Gaia Data Release 3 will contain more than a billion sources with positions, parallaxes, and proper motions. In addition, for hundreds of millions of stars, it will include low-resolution blue photometer (BP) and red photometer (RP) spectra. Obtained by dispersing light with prisms, these spectra have resolutions that are too low to allow us to measure individual spectral lines and bands. However, the combined BP/RP spectra can be used to estimate some stellar properties such as $\teff$, $\logg$, and $\mh$.}
{We investigate the feasibility of using the ExtraTrees algorithm to estimate the alpha element to iron abundance ratio $\afe$ from low-resolution BP/RP spectra.}
{To infer $\afe$ from the spectra, we created regression models using the ExtraTrees algorithm trained on two samples: a set of synthetic spectra and a set of observed spectra from stars that have known $\afe$ since they have also been observed using the High Efficiency and Resolution Multi-Element Spectrograph (HERMES) as part of the Galactic Archaeology with HERMES (GALAH) survey. We applied each model to the other sample and to a larger observed sample to assess the performance of the models. In addition, we used our models to analyse stars from the Gaia-Enceladus structure.}
{We find that the model trained on synthetic data has some ability to reconstruct $\afe$ from synthetic spectra, but little to none when used on observed spectra. The model trained on observed data reconstructs realistic $\afe$ from observed spectra, but only for cool stars ($\lessapprox 5000\,\unit{K}$) that have the same correlations as in the training sample between $\afe$ and other properties such as $\feh$.}
{Models using the ExtraTrees algorithm can be used to estimate $\afe$ from low-resolution BP/RP spectra of cool stars. However, they do this by exploiting correlations between $\afe$ and other parameters, rather than the causal effect of $\afe$ on the spectrum. Hence, they are unlikely to be useful in studies that attempt to distinguish stars that only differ in $\afe$.}
\keywords{Methods: statistical, Stars: abundances}

\begin{document}

\maketitle

\section{Introduction}
Since 2013, the Gaia telescope has been making observations of Milky Way stars~\citep{Gaia}. While the main objective of Gaia is astrometry, it also collects stellar spectra using the red photometer (RP), blue photometer (BP), and radial velocity spectrometer (RVS). The combined BP/RP spectra cover the wavelength range $3300$-$10500\,\unit{\AA}$ with a resolution between~13 and~85~\citep{XP}. The BP/RP spectra are observed for all targets, and a large subset will be released in Gaia Data Release~3 (GDR3). The BP/RP spectra can be used to estimate the stellar parameters effective temperature $\teff$, surface gravity $\logg$, and metallicity $\feh$ as well as the extinction along the line of sight $A_0$~\citep{gaia_spectrophotometry}. The RVS spectra can be used to estimate the same parameters, and also $\afe$ for cool stars. However, the RVS requires the stars to be relatively bright, so this will only be done for a few tens of million of stars out of the Gaia sample~\citep{RVS}. We aim to investigate whether the BP/RP spectra can also be used to estimate alpha abundances $\afe$ by using machine learning.

Estimating $\afe$ from BP/RP spectra is very much the type of problem that machine learning is intended for: we expect that the shape of the BP/RP spectra will depend on $\afe$ in some way that cannot easily be handled with conventional spectroscopic methods that measure line depths since the individual $\alpha$-sensitive spectral lines and bands cannot be resolved, but in large numbers they should still affect the continuum shape of the spectra. We also have known values for $\afe$, as well as other stellar parameters, for some subsets of the stars observed by Gaia from the surveys Galactic Archaeology with HERMES (GALAH) and Apache Point Observatory Galactic Evolution Experiment (APOGEE)~\citep{GALAH_buder, APOGEE}. Put in machine learning terms, we have features for a sample of objects in the form of BP/RP spectra, and we have known labels for a subset of those objects in the form of GALAH and APOGEE estimates of the $\afe$. We also have a theoretical reason to believe that labels correlate with features in some predictable way, but we lack a detailed model of what the relationship is.

Despite the argument outlined above, it could turn out that estimating $\afe$ from BP/RP spectra is impossible in practice due to degeneracies between $\afe$ and other parameters, or due to the effect of $\afe$ simply being swamped in noise. If it is possible, that leads to a second issue: even if the algorithm can estimate $\afe$ well on average, the basis for this estimate may or may not be something that will allow it to tell apart two stars that only differ by $\afe$. Ideally, we would want an algorithm that estimates $\afe$ based on the direct causal effect on the shape of the spectrum since this could be used to identify stars or populations of stars that have unusual $\afe$. On the other hand, an algorithm that estimates $\afe$ based on indirect correlations with other parameters -- at the most trivial, the Galactic trend of $\afe$ as a function of $\feh$ -- could not be used for this purpose, but may still be used cautiously in other contexts.

The accuracy with which $\afe$ can be estimated is expected to depend partly on the other stellar parameters. A priori, one would expect it to be easier to estimate for cooler stars, in particular for cool dwarf stars due to their prominent titanium oxide molecular bands which react quadratically to changes in $\alpha$-element abundance. However, even if machine learning can be reliably used on cool dwarfs, this is of limited use for Galactic population studies since Gaia can only observe those stars with high $\snr$ if they are very nearby.

Section~\ref{sec_extratrees} describes the machine learning algorithm used, the extremely randomized trees (ExtraTrees) algorithm. Section~\ref{sec_samples} describes the four samples of spectra that we use. Section~\ref{sec_train_synth} describes the results of training an ExtraTrees model on a sample of synthetic spectra. Section~\ref{sec_train_galah} describes the results of instead training an ExtraTrees model on a sample of spectra with parameters known through the GALAH survey. 
Section~\ref{sec_summary} summarises our results.

\section{ExtraTrees algorithm}\label{sec_extratrees}
For our machine learning algorithm we used the ExtraTrees algorithm, modified to allow simultaneous estimation of several outputs~\citep{ExtraTrees, multioutput}. The ExtraTrees algorithm is similar to the random forest algorithm, in that it is an ensemble learning method using decision trees as the base estimator. We used the implementation of the algorithm in the Python module scikit-learn~\citep{scikit-learn}. The algorithm has some advantages and some disadvantages, which lead us to prefer it over other algorithms.

One simple but important advantage is ease of use. It does not require lengthy fine-tuning of meta-parameters, and it is computationally simple enough to run very quickly. There are only three parameters that need to be defined when training the model. The first parameter is the maximum number of features $K$ to use to split a node when defining the decision trees~\citep[Sect. 2.1]{ExtraTrees}. We choose to always allow the full set of features to be used, since this is recommended for regression problems~\citep[Sect. 1.11.2.3]{scikit_manual}. The second parameter is the minimum sample size $n_{\text{min}}$ to split a node when defining the decision trees. When training on the synthetic sample (described in Sect.~\ref{sec_desc_synth}) we choose $n_{\text{min}} = 2$ since this typically gives the best results~\citep[Sect. 1.11.2.3]{scikit_manual}. When training on the GALAH sample (described in Sect.~\ref{sec_desc_GALAH}) we choose $n_{\text{min}} = 10$, since we otherwise ran into memory issues. The third parameter is the number of trees $n_\text{trees}$ to include in the ensemble. In principle, the algorithm will always perform better the higher $n_\text{trees}$ is, making the optimal value a trade-off with speed and memory. We choose $n_\text{trees} = 200$, since at this value we found that the uncertainties in the $\afe$ estimates due to the internal scatter among the base estimates were of the order of $0.01$-$0.03\,\unit{dex}$, making them safely negligible compared to other sources of uncertainty.

The algorithm has the second advantage that it can handle multi-output problems. This means that we could to some extent break the degeneracy between $\afe$ and other parameters by training the models to simultaneously estimate $\teff$, $\logg$, $\feh$, and $\afe$, instead of $\afe$ alone~\citep{multioutput}. When relevant, we show the estimates for the other parameters.

The algorithm has the third advantage that it is not very susceptible to overfit: many algorithms, such as neural nets or individual decision trees, are prone to learn features of the training sample that will not recur in any sample drawn from the same distribution, such as random noise. Ensemble learning methods are in general fairly robust to this issue, and the ExtraTrees algorithm was designed specifically to that end~\citep{ExtraTrees}.

One disadvantage of the algorithm is that being as simple as it is, it is likely to have slightly worse performance than a carefully calibrated version of some other algorithm, such as a neural networks. However, we believe that the algorithm is close enough to optimal that it is sufficient for the proof (or disproof) of concept that this study is intended to be. That is, if it is possible to do a measurement with the ExtraTrees algorithm, it may be possible to do it better with another algorithm -- but we do not believe that a measurement that is categorically impossible with the ExtraTrees algorithm will become possible with another algorithm.

A second disadvantage of the algorithm is that it is essentially incapable of extrapolating: as soon as it is used on data with features falling outside of the range represented in the training sample, the results are likely to be physically meaningless. To make our results easier to interpret we, where relevant, show the convex hull of the training sample in plots showing estimates made by the algorithm. In some cases we also filter our data to remove spectra that are not represented in the training sample.

The inability to extrapolate outside of the training sample also has the effect that the errors in estimates given by an ExtraTrees model are not necessarily symmetric: If a model is trained on some label covering the range $[x, y]$ and is then used to estimate labels for samples with true label values $x$ and $y$, then the former will always be overestimated and the latter will always be underestimated. This will be an issue for the synthetic sample described in Sect.~\ref{sec_desc_synth}, since it contains the discrete $\afe$ values~$0.0\,\unit{dex}$ and~$0.4\,\unit{dex}$.

\section{Spectral samples}\label{sec_samples}
We have four samples of BP/RP spectra: one sample of synthetic spectra that we dub the `synthetic sample' and describe in Sect.~\ref{sec_desc_synth}; One sample of observed BP/RP spectra with parameters known from the GALAH survey that we dub the `GALAH sample' and describe in Sect.~\ref{sec_desc_GALAH}; One sample of observed BP/RP spectra without known parameters that we dub the `Gaia sample' and describe in Sect.~\ref{sec_desc_Gaia}; One sample of observed BP/RP spectra that have parameters known from the APOGEE survey and are believed to be part of the Gaia-Enceladus structure, that we dub the `Gaia-Enceladus sample' and describe in Sect.~\ref{sec_desc_enc}.

Any of the three samples with known or estimated parameters can be used as a training sample, and the resulting model can then be used to estimate parameters for the same sample or any of the others. Section~\ref{sec_train_synth} describes the results of using the synthetic sample as training sample. Section~\ref{sec_train_galah} describes the results of using the GALAH sample as training sample.

\subsection{Synthetic sample}\label{sec_desc_synth}
We used the MARCS code to model a grid of synthetic spectra covering a range of stellar parameters and $\alpha$-abundances~\citep{marx}. We then extended the grid of spectra by adding an axis over the extinction parameter $A_0$ as defined in \citet[Sect. 2.2]{what_is_A0}. Finally we applied a model of the Gaia instrumental profile (Paolo Montegriffo, priv. comm.) This model has not yet been publicly documented, but may be released as part of the upcoming Gaia DR3.

\begin{figure}
\centering
\includegraphics[width=8.8cm]{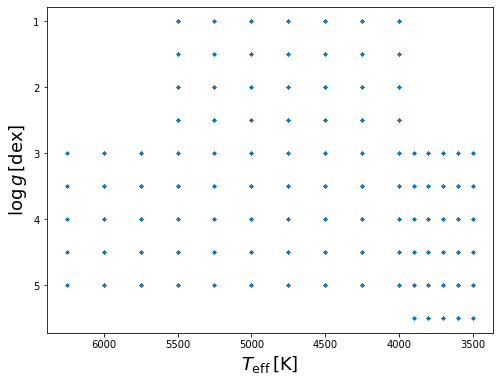}
\caption{Values of $\teff$ and $\logg$ of the grid of synthetic spectra. For each node, spectra with nine values of $\feh$, two values of $\afe$, and eleven values of $A_0$ have been calculated.}
\label{ill_grid}
\end{figure}

This sample consists of 21560 spectra in a piecewise regular grid, although a handful of spectra are missing due to specific MARCS models failing to converge. The sample contains three subsets of stars, which we dub `giants', `dwarfs', and `cool dwarfs'. The $\teff$-$\logg$ nodes are shown in Fig.~\ref{ill_grid}. The $\teff$ and $\logg$ values form three rectangles in $\teff$-$\logg$ space, one for each type of star. The giants cover $\teff$ between $4000$-$5500\,\unit{K}$ in steps of $250\,\unit{K}$ and $\logg$ between $1.0$-$3.0$ in steps of $0.5\,\unit{dex}$. The dwarfs cover $\teff$ between $4000$-$6250\,\unit{K}$ in steps of $250\,\unit{K}$ and $\logg$ between $3.0$-$5.0\,\unit{dex}$ in steps of $0.5\,\unit{dex}$. The cool dwarfs cover $\teff$ between $3500$ and $3900\,\unit{K}$ in steps of $100\,\unit{K}$, and $\logg$ of $3.0$-$5.5\,\unit{dex}$ in steps of $0.5\,\unit{dex}$. For each combination of $\teff$ and $\logg$ we calculate spectra with $\feh$ values from $-2.0\,\unit{dex}$ to $-1.0\,\unit{dex}$ in steps of $0.5\,\unit{dex}$, then to $0.25\,\unit{dex}$ in steps of $0.25\,\unit{dex}$; $\afe$ values of $0.0\,\unit{dex}$ and $0.4\,\unit{dex}$; $A_0$ values from $0.0$ to $1.0$ in steps of $0.1$.

\begin{figure}
\centering
\includegraphics[width=8.8cm]{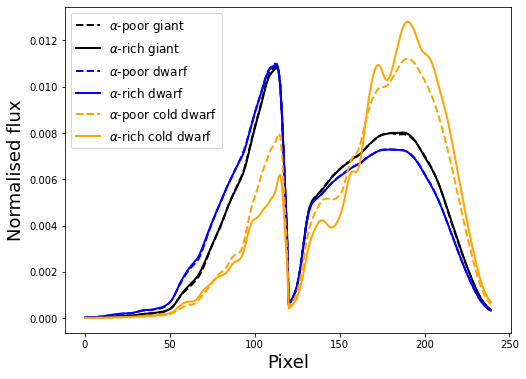}
\caption{Predicted BP/RP spectra, with $\afe$ of $0.0$ and $0.4\,\unit{dex}$, for representative spectra in the giant, dwarf, and cool dwarf sub-grids.}
\label{ill_synthetic_spectra}
\end{figure}

\begin{figure}
\centering
\includegraphics[width=8.8cm]{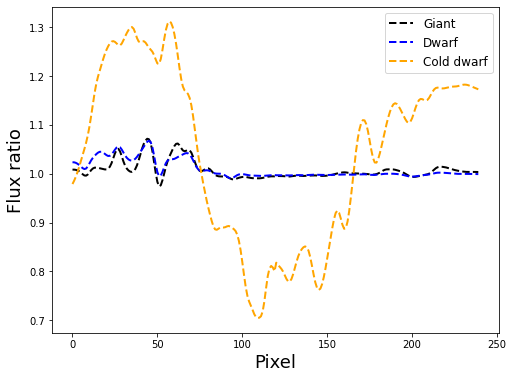}
\caption{Ratio between the fluxes in the synthetic spectra in Fig.~\ref{ill_synthetic_spectra}.}
\label{ill_synthetic_spectra_difference}
\end{figure}

Figure~\ref{ill_synthetic_spectra} shows synthetic BP/RP spectra for stars near the middle of each sub-sample, for $\afe = 0.0$ and $0.4\,\unit{dex}$. The giant spectra have $\teff = 4750\,\unit{K}$ and $\logg = 2.0\,\unit{dex}$. The dwarf spectra have $\teff = 5250\,\unit{K}$ and $\logg = 4.0\,\unit{dex}$. The cool dwarf spectra have $\teff = 3700\,\unit{K}$ and $\logg = 4.5\,\unit{dex}$. Figure~\ref{ill_synthetic_spectra_difference} shows the ratio in predicted flux between the $\alpha$-poor and $\alpha$-rich spectra. Based on this, one would expect performance to be best for the cool dwarfs, worse for giants and the worst for dwarfs.

For the scientific aim of this study, the coolest stars are probably of less interest than the rest of the sample, since their intrinsic faintness means that they are usually not used as tracers of Galactic trends. We originally included them because preliminary cross-validation with a smaller version of the synthetic sample indicated that the models had learned to use the edges of the grid to `cheat', breaking the degeneracies of $\afe$ with $\teff$ and $\logg$ by assuming that $\teff$ and $\logg$ have sharp cut-offs beyond which no stars exist. In principle this problem still exists -- the grid has to end somewhere -- but the grid should be large enough that the region of main scientific interest is unaffected.

\subsection{GALAH sample}\label{sec_desc_GALAH}
We constructed a sample of 188078 observed spectra of stars with stellar parameters that were estimated as part of Data Release~2 of the GALAH survey (GALAH DR2)~\citep{GALAH_dr2}. The full GALAH DR2 contains slightly below 350000 stars observed with the High Efficiency and Resolution Multi-Element Spectrograph (HERMES) at the Anglo-Australian Telescope (AAT). These have estimates of $\teff$, $\logg$, $\feh$, and $\afe$ which were made using the Cannon~\citep{cannon}. The Cannon was in turn trained on a representative subset of $10 605$ stars, for which stellar parameters were estimated using the spectral synthesis and fitting code Spectroscopy Made Easy (SME)~\citep{smeoriginal, smevolution}.

\begin{figure}
\centering
\includegraphics[width=8.8cm]{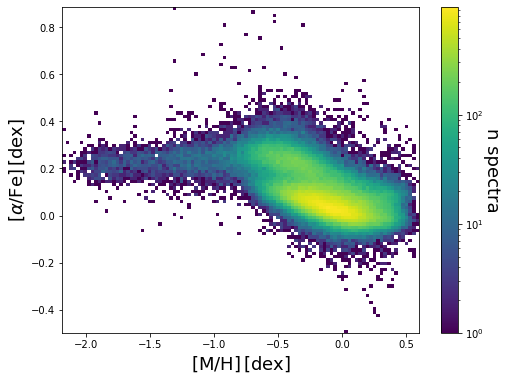}
\caption{Logarithmic-scale 2D histogram of the literature $\feh$ and $\afe$ for the GALAH sample.}
\label{ill_galah}
\end{figure}

\begin{figure}
\centering
\includegraphics[width=8.8cm]{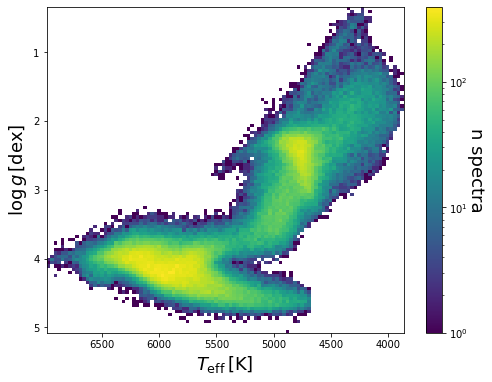}
\caption{Logarithmic-scale 2D histogram of the literature $\teff$ and $\logg$ for the GALAH sample.}
\label{ill_galah_teff_logg}
\end{figure}

We have applied the following selection criteria to construct our sample:
\begin{itemize}
    \item The GALAH uncertainty in $\teff$ is below $5\%$.
    \item The GALAH uncertainties in $\logg$ and $\feh$ are below $0.5\,\unit{dex}$.
    \item The BP/RP spectra must have at least five transits in both BP and RP.
    \item The BP/RP spectra have $\snr > 300$.
\item They must not have been flagged by The Cannon as possibly being unusual or having poor spectral reduction.
\item The $\snr$ in the GALAH green band must be larger than 25.
\item $\teff$ must be below $7000\,\unit{K}$, since stars hotter than that are underrepresented in the training sample of the Cannon.
\end{itemize}
Figure~\ref{ill_galah} shows the distribution over $\feh$ and $\afe$ of the sample. The two distinct peaks correspond to the metal-rich, $\alpha$-poor Thin Disk and the metal-poor $\alpha$-rich Thick Disk. Figure~\ref{ill_galah_teff_logg} shows the distribution over $\teff$ and $\logg$ for the sample. A model trained on this sample will not be applicable to spectra of any star that falls outside the parameter ranges represented in these plots.

\begin{figure}
\centering
\includegraphics[width=8.8cm]{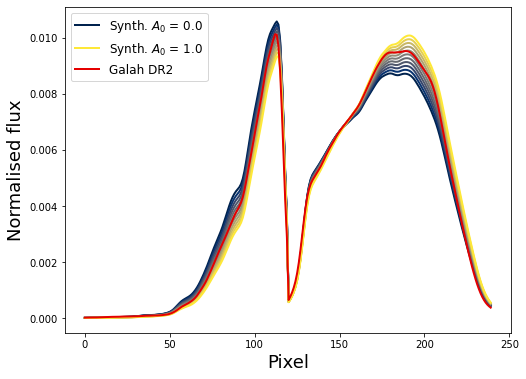}
\caption{Example spectrum from the GALAH sample shown in red. Shown below in blue to yellow are synthetic spectra with the same parameters, up to the step-length of the grid, and $A_0$ varying from~$0.0$ to~$1.0$.}
\label{ill_galah_synth_comp}
\end{figure}

Figure~\ref{ill_galah_synth_comp} shows in red a spectrum with literature parameters $\teff = 4324\,\unit{K}$, $\logg = 1.95\,\unit{dex}$, $\feh = 0.018\,\unit{dex}$, $\afe = 1.6 \cdot 10^{-3}\,\unit{dex}$. Plotted in yellow to blue are the corresponding synthetic spectra with parameters $4300\,\unit{K}$, $\logg = 2.0\,\unit{dex}$, $\feh = 0.0\,\unit{dex}$, $\afe = 0.0\,\unit{dex}$, covering $A_0$ from $0.0$ to $1.0$. The observed spectrum does not quite look like any of the synthetic spectra, or interpolation between them. This implies that the effects of the instrumental profile and extinction are not perfectly captured by our models, unless the star itself is atypical or has incorrectly estimated parameters.

\subsection{Gaia sample}\label{sec_desc_Gaia}
This sample is taken from the spectra currently being evaluated for inclusion in the upcoming GDR3, although some of them may end up being filtered out by the Gaia internal validation process. The spectra mostly do not have known stellar parameters. This means that they cannot be used to directly test how well a model performs, by comparing the parameter estimates to literature values. However, it can be used for indirect tests, by observing how well the parameter estimates replicate the Galactic structure.

\begin{figure}
\centering
\includegraphics[width=8.8cm]{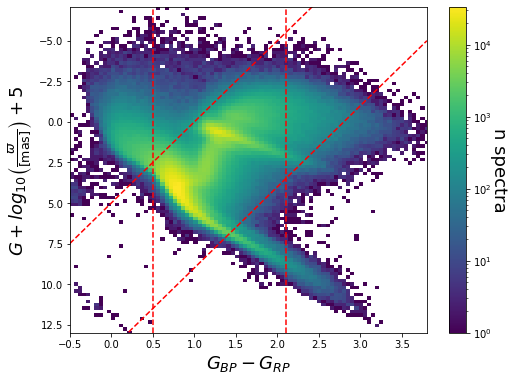}
\caption{Logarithmic-scale 2D histogram of the sum of the $G$ magnitude and the 10-logarithm of the Gaia parallax, and the BP-RP colour for the Gaia sample. Cuts given by Eqs.~\eqref{eq_cut_vertical} and~\eqref{eq_cut_diagonal} superimposed.}
\label{ill_gaia_sample}
\end{figure}

The spectra were originally selected based on the following criteria:
\begin{itemize}
\item They belong to the December 2019 version of the validation source table (VST) sample that was assembled by coordination unit 8 (CU8) of Gaia to validate the results of the Gaia astrophysical parameters inference system (APSIS).
\item The uncertainty in the parallax is better than $20\%$.
\item The BP/RP spectra have at least five transits in both BP and RP.
\end{itemize}
Since the ExtraTrees algorithm is incapable of extrapolating outside the range of the labels it has been trained on, we also apply the following cuts in magnitude to limit the sample to stars resembling those represented in the GALAH DR2 sample:
\begin{align}
&0.5 <  G_{\text{BP}} - G_{\text{RP}} < 2.1, \label{eq_cut_vertical} \\
& 0 < G + 5 \left( \log_{10} \left( \varpi / \left[ \unit{mas} \right] \right) + \left(  G_{\text{BP}} - G_{\text{RP}}  \right) \right) < 9, \label{eq_cut_diagonal}
\end{align}
where $G_{\text{BP}}$ is the integrated BP flux, $G_{\text{RP}}$ is the integrated RP flux, $G$ is the integrated $G$-filter flux and $\varpi$ is the Gaia parallax. Figure~\ref{ill_gaia_sample} shows this sample, with the cuts superimposed.

\subsection{Gaia-Enceladus sample}\label{sec_desc_enc}
Gaia-Enceladus is a kinematically and chemically distinct population of stars in the Galaxy. This structure was discovered in the kinematic data from Gaia DR2, together with the chemical data from the APOGEE survey. It is believed to be the remains of a dwarf galaxy slightly larger than the Small Magellanic Cloud, which was absorbed early in the history of the Galaxy~\citep{gaia-enceladus, gaia_dr2_summary, APOGEE}.

Gaia-Enceladus is chemically different from the rest of the Galaxy. It is more $\alpha$-rich than the Thin Disk and more metal-poor than either the Thick or the Thin Disk~\citep[Fig. 2]{gaia-enceladus} Because of this, it can be used as a test of what a model estimating $\afe$ is actually measuring. If a model is measuring the direct, causal effect of $\afe$, it should give estimates for Gaia-Enceladus that follow the true trend of $\afe$ as a function of $\feh$ in that structure. If the model is simply assuming that $\afe$ follows the Galactic trend, then that will be revealed by the estimates following the same trend for Gaia-Enceladus as for the rest of the Galaxy. If the model is using indirect correlations between $\afe$ and several other parameters, then the estimates can be expected to depart from those of the Galaxy without therefore following the true trend for Gaia-Enceladus.

\begin{figure}
\centering
\includegraphics[width=8.8cm]{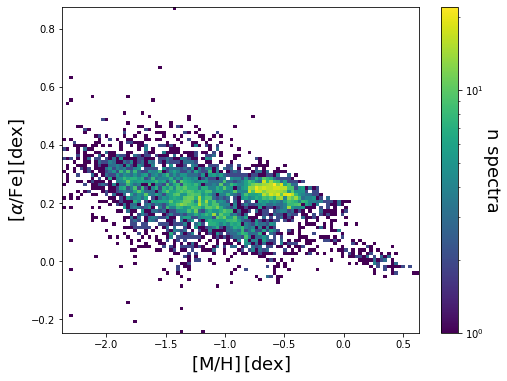}
\caption{Logarithmic-scale 2D histogram of the literature $\feh$ and $\afe$ for the Gaia-Enceladus sample.}
\label{ill_gaia-enceladus}
\end{figure}

We built a sample of 5783 stars (Carine Babusiaux, priv. comm.), selected as described in~\citet{gaia-enceladus}. For those stars, we collected BP/RP spectra which will be published in the upcoming GDR3. These spectra have parameters and abundances estimated as part of the APOGEE survey. Figure~\ref{ill_gaia-enceladus} shows the literature $\feh$ and $\afe$ of the sample.

\section{Training on synthetic sample}\label{sec_train_synth}
We made a regression model by training the ExtraTrees algorithm on the synthetic sample. In Sect.~\ref{sec_synth_crossval} we perform cross-validation to verify that the model can reconstruct its own training sample. In Sect.~\ref{sec_synth_categorisation} we investigate to what extent the reconstructed $\afe$ can be used to distinguish the $\alpha$-rich and $\alpha$-poor population in the training sample. In Sect.~\ref{sec_trainsynth_appgalah} we apply the model on the GALAH sample. In Sect.~\ref{sec_trainsynth_permutation} we calculate the permutation feature importance when applying the model to either sample. We do not show the results of applying the model to the Gaia and Gaia-Enceladus samples, since they lead to the same conclusions as the results for the GALAH sample.

\subsection{Cross-validation}\label{sec_synth_crossval}
We did ten-fold cross-validation on the training sample. This means that instead of training a model on the full sample, as we did in the rest of our analyses, we divided the sample into ten random subsets of approximately equal size. Then for each subset we trained a model using the other nine, and used that model to estimate the parameters for that subset. This essentially gave a best-case estimate of how well a model trained on the full sample can possibly be expected to perform, since it showed how well a sample with that statistical distribution over features and values can do on a sample with the exact same distribution.

By construction, the synthetic spectra are noise-free. This means that even if the ExtraTrees algorithm is robust to overfitting as such, there is a risk that models trained on these spectra may pick up on features in the data that would not be possible to distinguish in any observed spectrum. For our cross-validation, we therefore added Gaussian pixel noise to the spectra that were not currently used for training. To get realistic noise, we selected standard deviations from the uncertainty vector in that spectrum in the GALAH sample that had the most similar parameters. We took the most similar spectrum to be the one with the smallest distance $d$ defined as
\begin{equation}
d \equiv \sqrt{ \sum \left( \frac{ p_{\text{GALAH}} - p_{synth.}  }{\sigma_{\text{GALAH}}} \right)^2 }, \label{eq_dist}
\end{equation}
where $p_{\text{GALAH}}$ is the literature value of parameter $p$ for the spectrum in the GALAH sample, $p_{synth.}$ is the value of the same parameter in the synthetic sample and $\sigma_{\text{GALAH}}$ is the uncertainty in $p_{\text{GALAH}}$, and the summation runs over the parameters $\teff$, $\logg$, $\feh$, and $\afe$. As described in Appendix~\ref{app_noiseless_synth}, not including noise in the spectra that the model is applied to gave rise to unphysical artefacts in the estimates. On the other hand, including noise in the training spectra as well turned out to have a completely negligible effect.

\begin{figure}
\centering
\includegraphics[width=8.8cm]{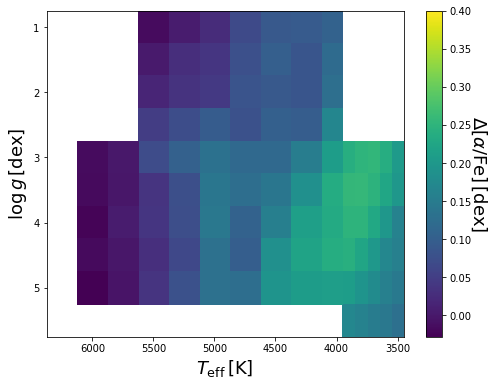}
\caption{Difference $\Delta \afe$ between the $\alpha$-rich and $\alpha$-poor stars in the synthetic sample, when performing cross-validation. For each $\teff$-$\logg$ node, the average is taken over $\feh$ and $A_0$.}
\label{ill_synthetic_alpha_difference}
\end{figure}

Figure~\ref{ill_synthetic_alpha_difference}, shows the difference between the estimated $\afe$ for the $\alpha$-rich and $\alpha$-poor spectra, at each $\teff$-$\logg$ node and averaged over $\feh$ and $A_0$. For the hottest stars, the spectra are practically indistinguishable. For some grid nodes the difference $\Delta \afe$ even drops slightly below zero, reaching as far as $-0.03\,\unit{dex}$.  As the temperature drops it rises above $0.2\,\unit{dex}$. This is considerably less than the actual difference, as expected due to the asymmetric errors described in Sect.~\ref{sec_extratrees}. Even so, some actual signal is clearly present.

\subsection{Categorisation}\label{sec_synth_categorisation}
To quantify the practical usefulness of the $\afe$ estimates shown in Fig.~\ref{ill_synthetic_alpha_difference}, we made a test that can serve as a toy model of a spectroscopic study. The assumption is that the spectroscopist is interested in distinguishing stellar populations based on the $\afe$ estimates, without being interested in the numeric estimates themselves.

\begin{figure}
\centering
\includegraphics[width=8.8cm]{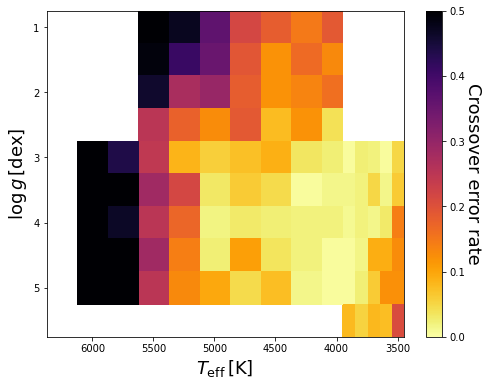}
\caption{Crossover error rate (CER) when attempting to distinguish $\alpha$-rich and $\alpha$-poor spectra when imposing a threshold~$\afe$ on the estimates derived during cross-validation of the synthetic sample. The black region around $6000\,\unit{K}$ has a CER near~$50\%$, indicating that the $\alpha$-rich and $\alpha$-poor spectra are practically indistinguishable. The pale yellow region around $4000\,\unit{K}$ has a CER below $10\%$, indicating that the $\alpha$-rich and $\alpha$-poor spectra can easily be told apart.}
\label{ill_synthetic_crossover_rate}
\end{figure}

In reality, each spectrum in the synthetic sample belongs to either a $\alpha$-poor population with $\afe = 0.0\,\unit{dex}$ or a $\alpha$-rich population with $\afe = 0.4\,\unit{dex}$. At each node in the $\teff$-$\logg$ grid we introduced some threshold value of $\afe$, and assigned all spectra below that to the $\alpha$-poor population and all spectra above to the $\alpha$-rich. We placed this threshold so that the fraction of mis-classified spectra is the same for both populations. (If the threshold were set low enough, all $\alpha$-rich and no $\alpha$-poor spectra would be correctly classified, and vice-versa). This crossover error rate (CER) is shown in Fig.~\ref{ill_synthetic_crossover_rate}. For the hottest stars the populations are almost indistinguishable, but as the temperature drops, the stars have a probability around $80$-$90\%$ of being correctly classified. The fact that the CER has a maximum around $4000\,\unit{K}$ rather than monotonously increasing as the temperature drops shows that the results are not an artefact of the grid edge. Based on this, it seemed that at least in an ideal case, the ExtraTrees algorithm could be useful in a study attempting to assign stars to populations, even if the numeric $\afe$ can be difficult to interpret.

\subsection{Application on GALAH sample}\label{sec_trainsynth_appgalah}
We used the model trained on the synthetic sample to estimate $\afe$ for the GALAH sample. This resulted in a root mean squared error (RMSE) of $0.2\,\unit{dex}$ when applied to the entire sample, and $0.1\,\unit{dex}$ when limited to giants.

To evaluate the results in more detail we looked at the normalised difference:
\begin{equation}
    \Delta^\text{norm}_p \equiv \frac{ p_\text{GALAH} - p_\text{ExtraTrees}}{ \sigma_\text{GALAH}}, \label{eq_normdiff}
\end{equation}
where $p_\text{GALAH}$ is the parameter estimate from GALAH, $p_\text{ExtraTrees}$ is the parameter estimate made with the Extratrees Algorithm trained on a non-overlapping tenth of the GALAH sample, and $\sigma_\text{GALAH}$ is the stated uncertainty in the GALAH estimate. The most ideal case possible would be for the GALAH estimates to be normally distributed around the true parameter values following normal distributions with standard deviations equal to the literature uncertainties, and for the estimates by the ExtraTrees model to always be exactly equal to the true parameter values. In that ideal case, Eq.~\eqref{eq_normdiff} would tend towards a standard normal distribution in the limit of large numbers of estimates. However, in Appendix~\ref{app_galah_underestimate} we find that at least for $\teff$ the GALAH survey has slightly overestimated errors, which causes our estimates to seem slightly more accurate than they actually are.

\begin{figure}
\centering
\includegraphics[width=8.8cm]{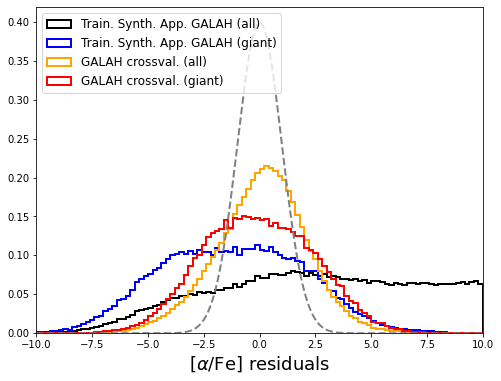}
\caption{Normalised difference, as defined in Eq.~\eqref{eq_normdiff}, between GALAH $\afe$ estimates and our estimates using the ExtraTrees algorithm.  If our performance was perfect, the distributions would in the limit of infinite data tend towards the standard normal distribution shown in grey.}
\label{ill_residuals}
\end{figure}

Figure~\ref{ill_residuals} shows histograms over $\Delta^\text{norm}_{\afe}$ for the full sample and for giant stars. It also shows the same for the cross-validation on the GALAH sample, discussed in Sect.~\ref{sec_galah_crossval}. While there is some ability to determine $\teff$, $\logg$, and $\feh$ (discussed in Appendix~\ref{app_galah_underestimate}), the values of $\afe$ are essentially random. It appears that the features of the synthetic spectra that the model has learned to use in the determination of $\afe$ are not robust enough to be usable in real-world spectra.

\subsection{Permutation feature importance}\label{sec_trainsynth_permutation}
While a model using the ExtraTrees algorithm is mostly opaque, there are a few tools for estimating the relative importance of different features. We used the permutation feature importance, which for a particular model and a sample with known labels attempts to estimate the importance of each feature by testing how much replacing that feature with a random value lowers the quality of the estimates. However, this method has the disadvantage that it can only be used on models that estimate a single parameter~\cite[Sect. 4.2]{scikit_manual}. Hence, we trained a model that only estimates $\afe$ and used it for this test, rather than the model simultaneously estimating $\teff$, $\logg$, $\feh$, and $\afe$ that we used in the rest of the article. The difference in performance between the models is small enough that conclusions based on the smaller model are likely to apply to the full model as well.

\begin{figure}
\centering
\includegraphics[width=8.8cm]{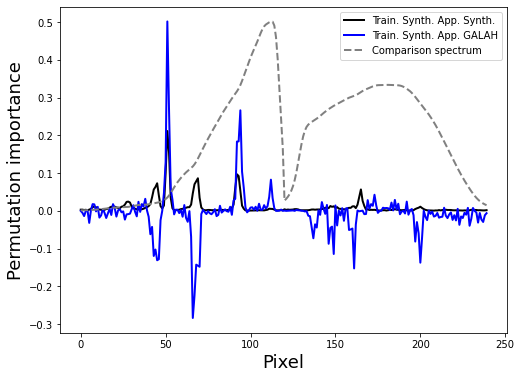}
\caption{Permutation feature importance when applying a model trained on the synthetic sample to the synthetic sample itself and to the GALAH sample. For comparison, the spectrum of the $\alpha$-rich dwarf star from Fig.~\ref{ill_synthetic_spectra} is shown in black, normalised to match the highest peak of the feature importances.}
\label{ill_permutation_feature_importance_synth}
\end{figure}

Figure~\ref{ill_permutation_feature_importance_synth} shows the permutation feature importance when applying the model trained on the synthetic sample on either the synthetic sample itself or the GALAH sample, together with a spectrum for comparison. There are two large peaks and one small peak where the permutation importance is higher when applying the model to the GALAH sample than when applying it to the synthetic sample. This shows that those pixels are genuinely important to getting useful estimates for the GALAH sample -- more so than in the synthetic sample. Over the rest of the spectrum the permutation importance is at best close to zero for the GALAH sample. In several places it is even negative, meaning that those pixels actively make the estimates worse.


\section{Training on GALAH sample}\label{sec_train_galah}
We made a second regression model by training the ExtraTrees algorithm on the GALAH sample. In Sect.~\ref{sec_galah_crossval} we perform cross-validation. In Sect.~\ref{sec_traingalah_appsynth} we use the model to estimate $\afe$ for the synthetic sample. In Sect.~\ref{sec_traingalah_permutation} we calculate the permutation feature importance when applying the model to either sample. In Sect.~\ref{sec_traingalah_appgaia} we use the model to estimate $\afe$ for the Gaia sample. In Sect.~\ref{sec_traingalah_appenc} we use the model to estimate $\afe$ for the Gaia-Enceladus sample.

\subsection{Cross validation}\label{sec_galah_crossval}
We performed ten-fold cross-validation analogously to Sect.~\ref{sec_synth_crossval}, except that we did not add synthetic noise. This resulted in a RMSE of $0.06\,\unit{dex}$ when applied to the entire sample, and $0.07\,\unit{dex}$ when limited to giants.

To evaluate the results in more detail, we again looked at the normalised difference as defined in Eq.~\eqref{eq_normdiff}. Figure~\ref{ill_residuals} shows histograms over the $\Delta^\text{norm}_{\afe}$ for the full sample and for giant stars. It also shows the same for a model trained on the synthetic sample, as discussed in Sect.~\ref{sec_trainsynth_appgalah}. Our $\afe$ estimates are less accurate than those of GALAH, but there is still clearly a signal. Our estimates for $\teff$, $\logg$, and $\feh$ (discussed in Appendix~\ref{app_galah_underestimate}) are of similar accuracy to GALAH.

\subsection{Application on synthetic sample}\label{sec_traingalah_appsynth}
\begin{figure}
\centering
\includegraphics[width=8.8cm]{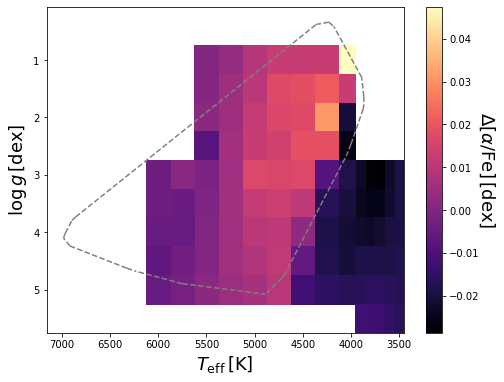}
\caption{Difference $\Delta \afe$ between estimated average $\afe$ for the $\alpha$-rich and $\alpha$-poor stars in the synthetic sample, after training on the GALAH sample. For each $\teff$-$\logg$ node, the average is taken over $\feh$ and $A_0$. Convex hull of training sample shown as dashed line. We note that if this plot had used the same colour scale as in Fig.~\ref{ill_synthetic_alpha_difference}, it would seem almost uniform to the eye.}
\label{ill_traingalah_appsynth}
\end{figure}

We used the model trained on the GALAH sample to estimate $\afe$ for the synthetic sample. Figure~\ref{ill_traingalah_appsynth} shows the difference in estimated $\afe$ for the $\alpha$-rich and $\alpha$-poor spectra, at each $\teff$-$\logg$ node and averaged over $\feh$ and $A_0$, analogously to Fig.~\ref{ill_synthetic_alpha_difference}. The convex hull of the training sample is overplotted. There is almost no difference between the two groups of spectra, with the best performance being just below $\Delta \afe = 0.05\,\unit{dex}$. This reveals that the information that the model has learned to pick out of the observed spectra is not actually present in the synthetic spectra. 

For the cool dwarfs the $\Delta \afe$ change sign, so that the estimated $\afe$ is actually slightly higher for the $\alpha$-poor spectra than the $\alpha$-rich. This reflects that those spectra fall outside of the range of features represented in the training sample: while the model correctly detects that the $\alpha$-rich and $\alpha$-poor spectra are different, it cannot use that information in any meaningful way. Otherwise we might have expected better performance for them, based on Fig.~\ref{ill_synthetic_crossover_rate}.

\subsection{Permutation feature importance}\label{sec_traingalah_permutation}
Similarly to the test on the model trained on the synthetic sample described in Sect.~\ref{sec_trainsynth_permutation}, we used the GALAH sample to train a model to only estimate $\afe$. We then calculated the permutation feature importance for this model with respect to the synthetic sample and to the GALAH sample itself.

\begin{figure}
\centering
\includegraphics[width=8.8cm]{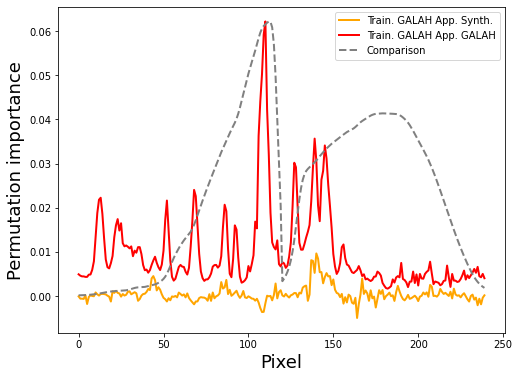}
\caption{Permutation feature importance when applying a model trained on the GALAH sample to the synthetic sample and to the GALAH sample itself. For comparison, the spectrum of the $\alpha$-rich dwarf star from Fig.~\ref{ill_synthetic_spectra} is shown in black, normalised to match the highest peak of the feature importances.}
\label{ill_permutation_feature_importance_GALAH}
\end{figure}

Figure~\ref{ill_permutation_feature_importance_GALAH} shows the permutation feature importance, together with a spectrum for comparison. When applying the model to the GALAH training sample, there are several peaks that show what pixels are important and useful in the estimates. When applying the same model to the synthetic sample there are no major peaks, reflecting the fact that the sample contains very little information that the model is able to use. When comparing to Fig.~\ref{ill_permutation_feature_importance_synth}, it is clear that the two models have learned to look at very different parts of the spectra.

\subsection{Application on Gaia sample}\label{sec_traingalah_appgaia}
\begin{figure}
\centering
\includegraphics[width=8.8cm]{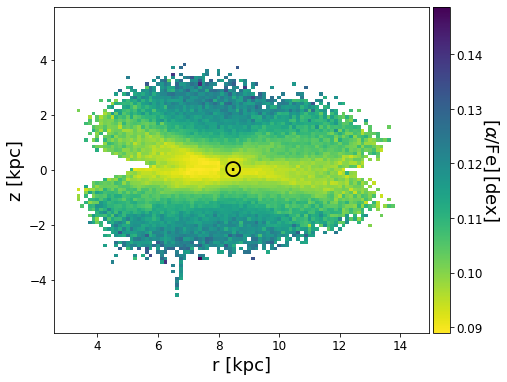}
\caption{Estimated $\afe$for the Gaia sample, using the GALAH sample as training sample, averaged over Galactocentric distance~$r$ and height~$z$ over Galactic plane. Position of the Sun shown for reference as a $\astrosun$.}
\label{ill_est_afe}
\includegraphics[width=8.8cm]{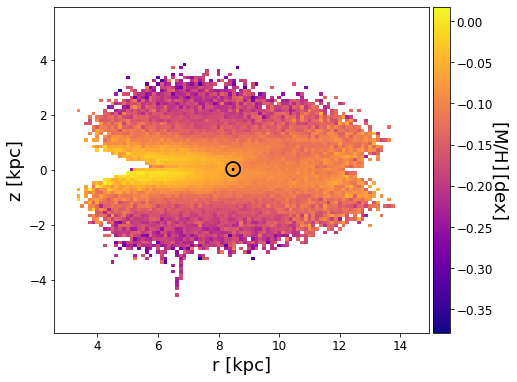}
\caption{Estimated $\feh$ for the Gaia sample. Interpretation otherwise the same as in Fig.~\ref{ill_est_afe}.}
\label{ill_est_feh}
\end{figure}

We used the model trained on the GALAH sample to estimate $\afe$ for the Gaia sample. We applied a cut at $\logg = 3.3$, since we found that the performance is different for giant stars. Figure~\ref{ill_est_afe} shows our average estimated $\afe$ as functions of Galactic position, for giant stars\footnote{Figures~\ref{ill_est_afe} and~\ref{ill_est_feh} have previously been released with minor differences as a Gaia image of the week~\citep{iotw}.}. This reveals a qualitatively realistic Galactic structure, with a $\afe$-poor Thin Disk and an $\afe$-rich Thick Disk. There is also flaring of the Disk at increasing radial distance. To verify this result, we show the corresponding estimated $\feh$ in Fig.~\ref{ill_est_feh}. This again shows a qualitatively realistic structure, with a $\feh$-rich Thin Disk and a $\feh$-poor Thick Disk.

This demonstrated that an ExtraTrees model trained on observed Gaia BP/RP spectra can make qualitatively realistic $\afe$ estimates for a different sample of observed Gaia BP/RP spectra. However, since it was unable to do so on the synthetic spectra, it appears to be doing this by using indirect correlations between $\afe$ and other stellar properties that have an effect on the spectrum, rather than the direct effect of $\afe$ on the spectrum. Such correlations exist in all observed samples, but by construction are missing from the synthetic sample.

This conclusion is slightly tentative, since it could in principle instead be that shortcomings in the modelling of the synthetic spectra made them unusable to a model trained on observed spectra. However, we do not believe so: While we know that there are shortcomings in our spectral synthesis, we do not expect this to result in information that is present in the observed spectra simply disappearing in the synthetic spectra. Rather, we would expect it to cause offsets in the estimates derived by a model trained on observed spectra, which is not what we see.

\subsection{Application on Gaia-Enceladus}\label{sec_traingalah_appenc}
Proceeding on the assumption that the model is using correlations between $\afe$ and other properties of the stars, we attempted to constrain what those properties are. At the most trivial, the model could simply have been using $\feh$ as a proxy for $\afe$ by using the Galactic trend of $\afe$ as a function of $\feh$. To test this, we used the model to estimate parameters for the Gaia-Enceladus sample. Gaia-Enceladus contains a significant fraction of stars which do not follow the Galactic trend in $\afe$.

\begin{figure}
\centering
\includegraphics[width=8.8cm]{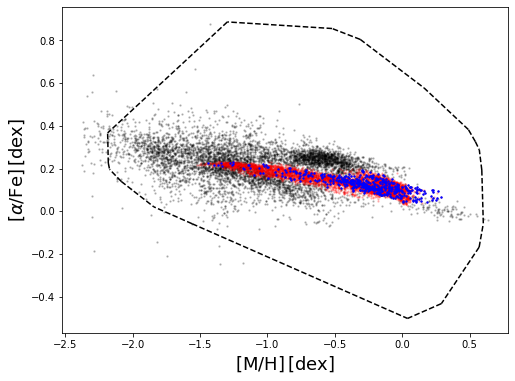}
\caption{Estimated parameters for Gaia-Enceladus using GALAH DR2 as training sample. $\alpha$-poor giant stars ($\afe < 0.1$, $\logg < 3.3$) are shown in blue, other stars in red. True parameters shown in the background in grey. Convex hull of training sample shown as dashed line.}
\label{ill_traingalah_appenc}
\end{figure}

Figure~\ref{ill_traingalah_appenc} shows the estimated parameters for the Gaia-Enceladus sample, marking the metal-poor giants in blue and other stars in red. The literature parameters are shown in grey, together with the convex hull of the training sample. It is apparent that the model is not able to reproduce the true distribution over $\afe$ and $\feh$ in Gaia-Enceladus.

\begin{figure}
\centering
\includegraphics[width=8.8cm]{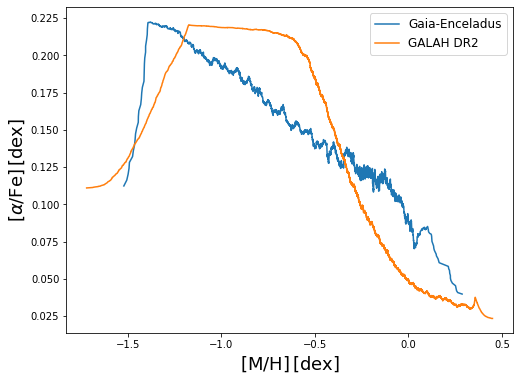}
\caption{Trend in $\afe$ as a function of $\feh$ for the GALAH DR2 and Gaia-Enceladus samples, using GALAH DR2 as training sample.}
\label{ill_traingalah_appenc_trends}
\end{figure}

Figure~\ref{ill_traingalah_appenc_trends} shows the trend of $\afe$ as function of $\feh$ for the GALAH DR2 and Gaia-Enceladus samples. The trends are estimated by plotting the $\feh$ and $\afe$ estimates for each star, and then smoothing with a kernel one 100th the width of the sample. The trends estimates are different, as they should be. Unfortunately, since we have seen that the models have difficulty discriminating between stars that only differ in $\afe$, it is unlikely that the models have genuinely captured the difference in $\afe$ between the samples. Instead, the difference between the estimated trends shows that the model does not simply use $\feh$ as a proxy for $\afe$, but takes other correlations into account as well.

\section{Summary and conclusion}\label{sec_summary}
We have attempted to find out if it is possible to use the ExtraTrees algorithm to estimate $\afe$ from Gaia BP/RP spectra. 
In our study we used four samples of spectra: The `synthetic sample', consisting of simulated spectra covering a grid of parameters; the `GALAH sample', consisting of observed spectra with parameters known from the GALAH survey; the `Gaia sample', consisting of observed spectra without known parameters; the `Gaia-Enceladus sample', consisting of observed spectra that are part of the Gaia-Enceladus structure and have known parameters from the APOGEE survey.

We first trained a model on the synthetic sample. When applied to synthetic spectra, the model could estimate $\afe$ with enough discrimination to allow distinguishing model populations of $\alpha$-rich and $\alpha$-poor stars.
We then applied the model to the GALAH sample and found that it was unable to estimate $\afe$ to any useful extent. Since models using the ExtraTrees algorithm are not very transparent, it was not possible to directly tell what information the model had learned to use from the synthetic sample, but this showed that that information is not actually present in observed spectra.

Next, we trained a model on the GALAH sample of observed spectra. We found that it was unable to estimate $\afe$ for the synthetic sample, but for the Gaia sample it did so well enough to reconstruct a realistic Galactic structure. Based on this we tentatively concluded that while the model could estimate $\afe$, it did so by using indirect correlations between $\afe$ and other properties that have an effect on the spectrum, rather than the direct, causal effect of $\afe$ on the spectrum.
We then applied the model to the Gaia-Enceladus sample, demonstrating that while the model does make use of indirect correlations, it does not merely treat $\feh$ as a proxy for $\afe$.

In the process of testing the GALAH sample we also found that, at least for the parameter $\teff$, the GALAH survey has slightly overestimated their own random errors. However, the discrepancy is likely small enough not to affect our conclusions.

In conclusion, we find that ExtraTrees models trained on observed spectra can make estimates of $\afe$ that are relatively close to the true values. However, those estimates are indirect -- based on the correlation between $\afe$ and other parameters including but not limited to $\feh$ -- rather than the direct causal effect of $\afe$ on the BP/RP spectra. This implies that this method cannot be used to distinguish stars that only differ in $\afe$. Finally, there are indications that cool dwarf stars may be an exception to this, but we cannot empirically verify this at present as our samples of observed Gaia BP/RP spectra do not cover that parameter range.

\begin{acknowledgements}
AG and AJK acknowledge support from the Swedish National Space Agency (SNSA). 
RA and MF contributions were funded in part by the DLR (German space agency) via grant 50 QG 1403. RS is supported by the Agenzia Spaziale Italiana (ASI) through contracts I/037/08/0, I/058/10/0, 2014-025-R.0, 2014-025-R.1.2015, and 2018-24-HH.0 to the Italian Istituto Nazionale di Astrofisica (INAF).
We thank Bengt Edvardsson for calculating synthetic spectra. We thank Carine Babusiaux for making the Gaia-Enceladus sample available to us. We thank Ulrike Heiter and Coryn Bailer-Jones for helpful feedback. We thank our colleagues from Gaia DPAC CU5, especially Dafydd Wyn Evans, Francesca De Angeli and Paolo Montegriffo, for their continuous support. 
This work has made use of data from the European Space Agency (ESA) mission Gaia (\url{http://www.cosmos.esa.int/gaia}), processed by the Gaia Data Processing and Analysis Consortium (DPAC, \url{http://www.cosmos.esa.int/web/gaia/dpac/consortium}). Funding for the DPAC has been provided by national institutions, in particular the institutions participating in the Gaia Multilateral Agreement.

\end{acknowledgements}

\bibliographystyle{aa}
\bibliography{bibliography}

\begin{thebibliography}{19}
\expandafter\ifx\csname natexlab\endcsname\relax\def\natexlab#1{#1}\fi

\bibitem[{Bailer-Jones(2011)}]{what_is_A0}
Bailer-Jones, C. A.~L. 2011, Monthly Notices of the Royal Astronomical Society,
  411, 435

\bibitem[{Buder {et~al.}(2018)Buder, Asplund, Duong, Kos, Lind, Ness, Sharma,
  Bland-Hawthorn, Casey, De Silva, D’Orazi, Freeman, Lewis, Lin, Martell,
  Schlesinger, Simpson, Zucker, Zwitter, Amarsi, Anguiano, Carollo, Casagrande,
  Čotar, Cottrell, Da Costa, Gao, Hayden, Horner, Ireland, Kafle, Munari,
  Nataf, Nordlander, Stello, Ting~(丁源森), Traven, Watson, Wittenmyer,
  Wyse, Yong, Zinn, Žerjal, \& collaboration}]{GALAH_dr2}
Buder, S., Asplund, M., Duong, L., {et~al.} 2018, Monthly Notices of the Royal
  Astronomical Society, 478, 4513

\bibitem[{{Buder, S.} {et~al.}(2019){Buder, S.}, {Lind, K.}, {Ness, M. K.},
  {Asplund, M.}, {Duong, L.}, {Lin, J.}, {Kos, J.}, {Casagrande, L.}, {Casey,
  A. R.}, {Bland-Hawthorn, J.}, {De Silva, G. M.}, {D\'{}Orazi, V.}, {Freeman,
  K. C.}, {Martell, S. L.}, {Schlesinger, K. J.}, {Sharma, S.}, {Simpson, J.
  D.}, {Zucker, D. B.}, {Zwitter, T.}, {Cotar, K.}, {Dotter, A.}, {Hayden, M.
  R.}, {Hyde, E. A.}, {Kafle, P. R.}, {Lewis, G. F.}, {Nataf, D. M.},
  {Nordlander, T.}, {Reid, W.}, {Rix, H.-W.}, {Sk\'ulad\'ottir, \'A.}, {Stello,
  D.}, {Ting, Y.-S.}, {Traven, G.}, {Wyse, R. F. G.}, \& {the GALAH
  collaboration}}]{GALAH_buder}
{Buder, S.}, {Lind, K.}, {Ness, M. K.}, {et~al.} 2019, A\&A, 624, A19

\bibitem[{{Carrasco} {et~al.}(2021){Carrasco}, {Weiler}, {Jordi}, {Fabricius},
  {De Angeli}, {Evans}, {van Leeuwen}, {Riello}, \& {Montegriffo}}]{XP}
{Carrasco}, J.~M., {Weiler}, M., {Jordi}, C., {et~al.} 2021, \aap, 652, A86

\bibitem[{Dumont {et~al.}(2009)Dumont, Marée, Wehenkel, \&
  Geurts}]{multioutput}
Dumont, M., Marée, R., Wehenkel, L., \& Geurts, P. 2009, in Proceedings of the
  Fourth International Conference on Computer Vision Theory and Applications,
  Vol.~2, 196--203

\bibitem[{{Gaia Collaboration} {et~al.}(2018){Gaia Collaboration}, {Brown, A.
  G. A.}, {Vallenari, A.}, {Prusti, T.}, {de Bruijne, J. H. J.}, {Babusiaux,
  C.}, {Bailer-Jones, C. A. L.}, {Biermann, M.}, {Evans, D. W.}, {Eyer, L.},
  {Jansen, F.}, {Jordi, C.}, {Klioner, S. A.}, {Lammers, U.}, {Lindegren, L.},
  {Luri, X.}, {Mignard, F.}, {Panem, C.}, {Pourbaix, D.}, {Randich, S.},
  {Sartoretti, P.}, {Siddiqui, H. I.}, {Soubiran, C.}, {van Leeuwen, F.},
  {Walton, N. A.}, {Arenou, F.}, {Bastian, U.}, {Cropper, M.}, {Drimmel, R.},
  {Katz, D.}, {Lattanzi, M. G.}, {Bakker, J.}, {Cacciari, C.}, {Casta\~neda,
  J.}, {Chaoul, L.}, {Cheek, N.}, {De Angeli, F.}, {Fabricius, C.}, {Guerra,
  R.}, {Holl, B.}, {Masana, E.}, {Messineo, R.}, {Mowlavi, N.}, {Nienartowicz,
  K.}, {Panuzzo, P.}, {Portell, J.}, {Riello, M.}, {Seabroke, G. M.}, {Tanga,
  P.}, {Th\'evenin, F.}, {Gracia-Abril, G.}, {Comoretto, G.},
  {Garcia-Reinaldos, M.}, {Teyssier, D.}, {Altmann, M.}, {Andrae, R.}, {Audard,
  M.}, {Bellas-Velidis, I.}, {Benson, K.}, {Berthier, J.}, {Blomme, R.},
  {Burgess, P.}, {Busso, G.}, {Carry, B.}, {Cellino, A.}, {Clementini, G.},
  {Clotet, M.}, {Creevey, O.}, {Davidson, M.}, {De Ridder, J.}, {Delchambre,
  L.}, {Dell\'{}Oro, A.}, {Ducourant, C.}, {Fern\'andez-Hern\'andez, J.},
  {Fouesneau, M.}, {Fr\'emat, Y.}, {Galluccio, L.}, {Garc\'{\i}a-Torres, M.},
  {Gonz\'alez-N\'u\~nez, J.}, {Gonz\'alez-Vidal, J. J.}, {Gosset, E.}, {Guy, L.
  P.}, {Halbwachs, J.-L.}, {Hambly, N. C.}, {Harrison, D. L.}, {Hern\'andez,
  J.}, {Hestroffer, D.}, {Hodgkin, S. T.}, {Hutton, A.}, {Jasniewicz, G.},
  {Jean-Antoine-Piccolo, A.}, {Jordan, S.}, {Korn, A. J.}, {Krone-Martins, A.},
  {Lanzafame, A. C.}, {Lebzelter, T.}, {L\"offler, W.}, {Manteiga, M.},
  {Marrese, P. M.}, {Mart\'{\i}n-Fleitas, J. M.}, {Moitinho, A.}, {Mora, A.},
  {Muinonen, K.}, {Osinde, J.}, {Pancino, E.}, {Pauwels, T.}, {Petit, J.-M.},
  {Recio-Blanco, A.}, {Richards, P. J.}, {Rimoldini, L.}, {Robin, A. C.},
  {Sarro, L. M.}, {Siopis, C.}, {Smith, M.}, {Sozzetti, A.}, {S\"uveges, M.},
  {Torra, J.}, {van Reeven, W.}, {Abbas, U.}, {Abreu Aramburu, A.}, {Accart,
  S.}, {Aerts, C.}, {Altavilla, G.}, {\'Alvarez, M. A.}, {Alvarez, R.}, {Alves,
  J.}, {Anderson, R. I.}, {Andrei, A. H.}, {Anglada Varela, E.}, {Antiche, E.},
  {Antoja, T.}, {Arcay, B.}, {Astraatmadja, T. L.}, {Bach, N.}, {Baker, S. G.},
  {Balaguer-N\'u\~nez, L.}, {Balm, P.}, {Barache, C.}, {Barata, C.}, {Barbato,
  D.}, {Barblan, F.}, {Barklem, P. S.}, {Barrado, D.}, {Barros, M.}, {Barstow,
  M. A.}, {Bartholom\'e Mu\~noz, S.}, {Bassilana, J.-L.}, {Becciani, U.},
  {Bellazzini, M.}, {Berihuete, A.}, {Bertone, S.}, {Bianchi, L.}, {Bienaym\'e,
  O.}, {Blanco-Cuaresma, S.}, {Boch, T.}, {Boeche, C.}, {Bombrun, A.},
  {Borrachero, R.}, {Bossini, D.}, {Bouquillon, S.}, {Bourda, G.}, {Bragaglia,
  A.}, {Bramante, L.}, {Breddels, M. A.}, {Bressan, A.}, {Brouillet, N.},
  {Br\"usemeister, T.}, {Brugaletta, E.}, {Bucciarelli, B.}, {Burlacu, A.},
  {Busonero, D.}, {Butkevich, A. G.}, {Buzzi, R.}, {Caffau, E.}, {Cancelliere,
  R.}, {Cannizzaro, G.}, {Cantat-Gaudin, T.}, {Carballo, R.}, {Carlucci, T.},
  {Carrasco, J. M.}, {Casamiquela, L.}, {Castellani, M.}, {Castro-Ginard, A.},
  {Charlot, P.}, {Chemin, L.}, {Chiavassa, A.}, {Cocozza, G.}, {Costigan, G.},
  {Cowell, S.}, {Crifo, F.}, {Crosta, M.}, {Crowley, C.}, {Cuypers+, J.},
  {Dafonte, C.}, {Damerdji, Y.}, {Dapergolas, A.}, {David, P.}, {David, M.},
  {de Laverny, P.}, {De Luise, F.}, {De March, R.}, {de Martino, D.}, {de
  Souza, R.}, {de Torres, A.}, {Debosscher, J.}, {del Pozo, E.}, {Delbo, M.},
  {Delgado, A.}, {Delgado, H. E.}, {Di Matteo, P.}, {Diakite, S.}, {Diener,
  C.}, {Distefano, E.}, {Dolding, C.}, {Drazinos, P.}, {Dur\'an, J.},
  {Edvardsson, B.}, {Enke, H.}, {Eriksson, K.}, {Esquej, P.}, {Eynard Bontemps,
  G.}, {Fabre, C.}, {Fabrizio, M.}, {Faigler, S.}, {Falc\~ao, A. J.}, {Farr\`as
  Casas, M.}, {Federici, L.}, {Fedorets, G.}, {Fernique, P.}, {Figueras, F.},
  {Filippi, F.}, {Findeisen, K.}, {Fonti, A.}, {Fraile, E.}, {Fraser, M.},
  {Fr\'ezouls, B.}, {Gai, M.}, {Galleti, S.}, {Garabato, D.},
  {Garc\'{\i}a-Sedano, F.}, {Garofalo, A.}, {Garralda, N.}, {Gavel, A.},
  {Gavras, P.}, {Gerssen, J.}, {Geyer, R.}, {Giacobbe, P.}, {Gilmore, G.},
  {Girona, S.}, {Giuffrida, G.}, {Glass, F.}, {Gomes, M.}, {Granvik, M.},
  {Gueguen, A.}, {Guerrier, A.}, {Guiraud, J.}, {Guti\'errez-S\'anchez, R.},
  {Haigron, R.}, {Hatzidimitriou, D.}, {Hauser, M.}, {Haywood, M.}, {Heiter,
  U.}, {Helmi, A.}, {Heu, J.}, {Hilger, T.}, {Hobbs, D.}, {Hofmann, W.},
  {Holland, G.}, {Huckle, H. E.}, {Hypki, A.}, {Icardi, V.}, {Jan\ss{}en, K.},
  {Jevardat de Fombelle, G.}, {Jonker, P. G.}, {Juh\'asz, \'A. L.}, {Julbe,
  F.}, {Karampelas, A.}, {Kewley, A.}, {Klar, J.}, {Kochoska, A.}, {Kohley,
  R.}, {Kolenberg, K.}, {Kontizas, M.}, {Kontizas, E.}, {Koposov, S. E.},
  {Kordopatis, G.}, {Kostrzewa-Rutkowska, Z.}, {Koubsky, P.}, {Lambert, S.},
  {Lanza, A. F.}, {Lasne, Y.}, {Lavigne, J.-B.}, {Le Fustec, Y.}, {Le
  Poncin-Lafitte, C.}, {Lebreton, Y.}, {Leccia, S.}, {Leclerc, N.},
  {Lecoeur-Taibi, I.}, {Lenhardt, H.}, {Leroux, F.}, {Liao, S.}, {Licata, E.},
  {Lindstr\o{}m, H. E. P.}, {Lister, T. A.}, {Livanou, E.}, {Lobel, A.},
  {L\'opez, M.}, {Managau, S.}, {Mann, R. G.}, {Mantelet, G.}, {Marchal, O.},
  {Marchant, J. M.}, {Marconi, M.}, {Marinoni, S.}, {Marschalk\'o, G.},
  {Marshall, D. J.}, {Martino, M.}, {Marton, G.}, {Mary, N.}, {Massari, D.},
  {Matijevic, G.}, {Mazeh, T.}, {McMillan, P. J.}, {Messina, S.}, {Michalik,
  D.}, {Millar, N. R.}, {Molina, D.}, {Molinaro, R.}, {Moln\'ar, L.},
  {Montegriffo, P.}, {Mor, R.}, {Morbidelli, R.}, {Morel, T.}, {Morris, D.},
  {Mulone, A. F.}, {Muraveva, T.}, {Musella, I.}, {Nelemans, G.}, {Nicastro,
  L.}, {Noval, L.}, {O\'{}Mullane, W.}, {Ord\'enovic, C.}, {Ord\'o\~nez-Blanco,
  D.}, {Osborne, P.}, {Pagani, C.}, {Pagano, I.}, {Pailler, F.}, {Palacin, H.},
  {Palaversa, L.}, {Panahi, A.}, {Pawlak, M.}, {Piersimoni, A. M.}, {Pineau,
  F.-X.}, {Plachy, E.}, {Plum, G.}, {Poggio, E.}, {Poujoulet, E.}, {Prsa, A.},
  {Pulone, L.}, {Racero, E.}, {Ragaini, S.}, {Rambaux, N.}, {Ramos-Lerate, M.},
  {Regibo, S.}, {Reyl\'e, C.}, {Riclet, F.}, {Ripepi, V.}, {Riva, A.}, {Rivard,
  A.}, {Rixon, G.}, {Roegiers, T.}, {Roelens, M.}, {Romero-G\'omez, M.},
  {Rowell, N.}, {Royer, F.}, {Ruiz-Dern, L.}, {Sadowski, G.}, {Sagrist\`a
  Sell\'es, T.}, {Sahlmann, J.}, {Salgado, J.}, {Salguero, E.}, {Sanna, N.},
  {Santana-Ros, T.}, {Sarasso, M.}, {Savietto, H.}, {Schultheis, M.}, {Sciacca,
  E.}, {Segol, M.}, {Segovia, J. C.}, {S\'egransan, D.}, {Shih, I-C.},
  {Siltala, L.}, {Silva, A. F.}, {Smart, R. L.}, {Smith, K. W.}, {Solano, E.},
  {Solitro, F.}, {Sordo, R.}, {Soria Nieto, S.}, {Souchay, J.}, {Spagna, A.},
  {Spoto, F.}, {Stampa, U.}, {Steele, I. A.}, {Steidelm\"uller, H.},
  {Stephenson, C. A.}, {Stoev, H.}, {Suess, F. F.}, {Surdej, J.}, {Szabados,
  L.}, {Szegedi-Elek, E.}, {Tapiador, D.}, {Taris, F.}, {Tauran, G.}, {Taylor,
  M. B.}, {Teixeira, R.}, {Terrett, D.}, {Teyssandier, P.}, {Thuillot, W.},
  {Titarenko, A.}, {Torra Clotet, F.}, {Turon, C.}, {Ulla, A.}, {Utrilla, E.},
  {Uzzi, S.}, {Vaillant, M.}, {Valentini, G.}, {Valette, V.}, {van Elteren,
  A.}, {Van Hemelryck, E.}, {van Leeuwen, M.}, {Vaschetto, M.}, {Vecchiato,
  A.}, {Veljanoski, J.}, {Viala, Y.}, {Vicente, D.}, {Vogt, S.}, {von Essen,
  C.}, {Voss, H.}, {Votruba, V.}, {Voutsinas, S.}, {Walmsley, G.}, {Weiler,
  M.}, {Wertz, O.}, {Wevers, T.}, {Wyrzykowski, L.}, {Yoldas, A.}, {Zerjal,
  M.}, {Ziaeepour, H.}, {Zorec, J.}, {Zschocke, S.}, {Zucker, S.}, {Zurbach,
  C.}, \& {Zwitter, T.}}]{gaia_dr2_summary}
{Gaia Collaboration}, {Brown, A. G. A.}, {Vallenari, A.}, {et~al.} 2018, A\&A,
  616, A1

\bibitem[{{Gaia Collaboration} {et~al.}(2016){Gaia Collaboration}, {Prusti},
  {de Bruijne}, {Brown}, {Vallenari}, {Babusiaux}, {Bailer-Jones}, {Bastian},
  {Biermann}, {Evans}, {Eyer}, {Jansen}, {Jordi}, {Klioner}, {Lammers},
  {Lindegren}, {Luri}, {Mignard}, {Milligan}, {Panem}, {Poinsignon},
  {Pourbaix}, {Randich}, {Sarri}, {Sartoretti}, {Siddiqui}, {Soubiran},
  {Valette}, {van Leeuwen}, {Walton}, {Aerts}, {Arenou}, {Cropper}, {Drimmel},
  {H{\o}g}, {Katz}, {Lattanzi}, {O'Mullane}, {Grebel}, {Holland}, {Huc},
  {Passot}, {Bramante}, {Cacciari}, {Casta{\~n}eda}, {Chaoul}, {Cheek}, {De
  Angeli}, {Fabricius}, {Guerra}, {Hern{\'a}ndez}, {Jean-Antoine-Piccolo},
  {Masana}, {Messineo}, {Mowlavi}, {Nienartowicz}, {Ord{\'o}{\~n}ez-Blanco},
  {Panuzzo}, {Portell}, {Richards}, {Riello}, {Seabroke}, {Tanga},
  {Th{\'e}venin}, {Torra}, {Els}, {Gracia-Abril}, {Comoretto},
  {Garcia-Reinaldos}, {Lock}, {Mercier}, {Altmann}, {Andrae}, {Astraatmadja},
  {Bellas-Velidis}, {Benson}, {Berthier}, {Blomme}, {Busso}, {Carry},
  {Cellino}, {Clementini}, {Cowell}, {Creevey}, {Cuypers}, {Davidson}, {De
  Ridder}, {de Torres}, {Delchambre}, {Dell'Oro}, {Ducourant}, {Fr{\'e}mat},
  {Garc{\'\i}a-Torres}, {Gosset}, {Halbwachs}, {Hambly}, {Harrison}, {Hauser},
  {Hestroffer}, {Hodgkin}, {Huckle}, {Hutton}, {Jasniewicz}, {Jordan},
  {Kontizas}, {Korn}, {Lanzafame}, {Manteiga}, {Moitinho}, {Muinonen},
  {Osinde}, {Pancino}, {Pauwels}, {Petit}, {Recio-Blanco}, {Robin}, {Sarro},
  {Siopis}, {Smith}, {Smith}, {Sozzetti}, {Thuillot}, {van Reeven}, {Viala},
  {Abbas}, {Abreu Aramburu}, {Accart}, {Aguado}, {Allan}, {Allasia},
  {Altavilla}, {{\'A}lvarez}, {Alves}, {Anderson}, {Andrei}, {Anglada Varela},
  {Antiche}, {Antoja}, {Ant{\'o}n}, {Arcay}, {Atzei}, {Ayache}, {Bach},
  {Baker}, {Balaguer-N{\'u}{\~n}ez}, {Barache}, {Barata}, {Barbier}, {Barblan},
  {Baroni}, {Barrado y Navascu{\'e}s}, {Barros}, {Barstow}, {Becciani},
  {Bellazzini}, {Bellei}, {Bello Garc{\'\i}a}, {Belokurov}, {Bendjoya},
  {Berihuete}, {Bianchi}, {Bienaym{\'e}}, {Billebaud}, {Blagorodnova},
  {Blanco-Cuaresma}, {Boch}, {Bombrun}, {Borrachero}, {Bouquillon}, {Bourda},
  {Bouy}, {Bragaglia}, {Breddels}, {Brouillet}, {Br{\"u}semeister},
  {Bucciarelli}, {Budnik}, {Burgess}, {Burgon}, {Burlacu}, {Busonero}, {Buzzi},
  {Caffau}, {Cambras}, {Campbell}, {Cancelliere}, {Cantat-Gaudin}, {Carlucci},
  {Carrasco}, {Castellani}, {Charlot}, {Charnas}, {Charvet}, {Chassat},
  {Chiavassa}, {Clotet}, {Cocozza}, {Collins}, {Collins}, {Costigan}, {Crifo},
  {Cross}, {Crosta}, {Crowley}, {Dafonte}, {Damerdji}, {Dapergolas}, {David},
  {David}, {De Cat}, {de Felice}, {de Laverny}, {De Luise}, {De March}, {de
  Martino}, {de Souza}, {Debosscher}, {del Pozo}, {Delbo}, {Delgado},
  {Delgado}, {di Marco}, {Di Matteo}, {Diakite}, {Distefano}, {Dolding}, {Dos
  Anjos}, {Drazinos}, {Dur{\'a}n}, {Dzigan}, {Ecale}, {Edvardsson}, {Enke},
  {Erdmann}, {Escolar}, {Espina}, {Evans}, {Eynard Bontemps}, {Fabre},
  {Fabrizio}, {Faigler}, {Falc{\~a}o}, {Farr{\`a}s Casas}, {Faye}, {Federici},
  {Fedorets}, {Fern{\'a}ndez-Hern{\'a}ndez}, {Fernique}, {Fienga}, {Figueras},
  {Filippi}, {Findeisen}, {Fonti}, {Fouesneau}, {Fraile}, {Fraser}, {Fuchs},
  {Furnell}, {Gai}, {Galleti}, {Galluccio}, {Garabato}, {Garc{\'\i}a-Sedano},
  {Gar{\'e}}, {Garofalo}, {Garralda}, {Gavras}, {Gerssen}, {Geyer}, {Gilmore},
  {Girona}, {Giuffrida}, {Gomes}, {Gonz{\'a}lez-Marcos},
  {Gonz{\'a}lez-N{\'u}{\~n}ez}, {Gonz{\'a}lez-Vidal}, {Granvik}, {Guerrier},
  {Guillout}, {Guiraud}, {G{\'u}rpide}, {Guti{\'e}rrez-S{\'a}nchez}, {Guy},
  {Haigron}, {Hatzidimitriou}, {Haywood}, {Heiter}, {Helmi}, {Hobbs},
  {Hofmann}, {Holl}, {Holland}, {Hunt}, {Hypki}, {Icardi}, {Irwin}, {Jevardat
  de Fombelle}, {Jofr{\'e}}, {Jonker}, {Jorissen}, {Julbe}, {Karampelas},
  {Kochoska}, {Kohley}, {Kolenberg}, {Kontizas}, {Koposov}, {Kordopatis},
  {Koubsky}, {Kowalczyk}, {Krone-Martins}, {Kudryashova}, {Kull}, {Bachchan},
  {Lacoste-Seris}, {Lanza}, {Lavigne}, {Le Poncin-Lafitte}, {Lebreton},
  {Lebzelter}, {Leccia}, {Leclerc}, {Lecoeur-Taibi}, {Lemaitre}, {Lenhardt},
  {Leroux}, {Liao}, {Licata}, {Lindstr{\o}m}, {Lister}, {Livanou}, {Lobel},
  {L{\"o}ffler}, {L{\'o}pez}, {Lopez-Lozano}, {Lorenz}, {Loureiro},
  {MacDonald}, {Magalh{\~a}es Fernandes}, {Managau}, {Mann}, {Mantelet},
  {Marchal}, {Marchant}, {Marconi}, {Marie}, {Marinoni}, {Marrese},
  {Marschalk{\'o}}, {Marshall}, {Mart{\'\i}n-Fleitas}, {Martino}, {Mary},
  {Matijevi{\v{c}}}, {Mazeh}, {McMillan}, {Messina}, {Mestre}, {Michalik},
  {Millar}, {Miranda}, {Molina}, {Molinaro}, {Molinaro}, {Moln{\'a}r},
  {Moniez}, {Montegriffo}, {Monteiro}, {Mor}, {Mora}, {Morbidelli}, {Morel},
  {Morgenthaler}, {Morley}, {Morris}, {Mulone}, {Muraveva}, {Musella},
  {Narbonne}, {Nelemans}, {Nicastro}, {Noval}, {Ord{\'e}novic},
  {Ordieres-Mer{\'e}}, {Osborne}, {Pagani}, {Pagano}, {Pailler}, {Palacin},
  {Palaversa}, {Parsons}, {Paulsen}, {Pecoraro}, {Pedrosa}, {Pentik{\"a}inen},
  {Pereira}, {Pichon}, {Piersimoni}, {Pineau}, {Plachy}, {Plum}, {Poujoulet},
  {Pr{\v{s}}a}, {Pulone}, {Ragaini}, {Rago}, {Rambaux}, {Ramos-Lerate},
  {Ranalli}, {Rauw}, {Read}, {Regibo}, {Renk}, {Reyl{\'e}}, {Ribeiro},
  {Rimoldini}, {Ripepi}, {Riva}, {Rixon}, {Roelens}, {Romero-G{\'o}mez},
  {Rowell}, {Royer}, {Rudolph}, {Ruiz-Dern}, {Sadowski}, {Sagrist{\`a}
  Sell{\'e}s}, {Sahlmann}, {Salgado}, {Salguero}, {Sarasso}, {Savietto},
  {Schnorhk}, {Schultheis}, {Sciacca}, {Segol}, {Segovia}, {Segransan},
  {Serpell}, {Shih}, {Smareglia}, {Smart}, {Smith}, {Solano}, {Solitro},
  {Sordo}, {Soria Nieto}, {Souchay}, {Spagna}, {Spoto}, {Stampa}, {Steele},
  {Steidelm{\"u}ller}, {Stephenson}, {Stoev}, {Suess}, {S{\"u}veges}, {Surdej},
  {Szabados}, {Szegedi-Elek}, {Tapiador}, {Taris}, {Tauran}, {Taylor},
  {Teixeira}, {Terrett}, {Tingley}, {Trager}, {Turon}, {Ulla}, {Utrilla},
  {Valentini}, {van Elteren}, {Van Hemelryck}, {van Leeuwen}, {Varadi},
  {Vecchiato}, {Veljanoski}, {Via}, {Vicente}, {Vogt}, {Voss}, {Votruba},
  {Voutsinas}, {Walmsley}, {Weiler}, {Weingrill}, {Werner}, {Wevers},
  {Whitehead}, {Wyrzykowski}, {Yoldas}, {{\v{Z}}erjal}, {Zucker}, {Zurbach},
  {Zwitter}, {Alecu}, {Allen}, {Allende Prieto}, {Amorim},
  {Anglada-Escud{\'e}}, {Arsenijevic}, {Azaz}, {Balm}, {Beck}, {Bernstein},
  {Bigot}, {Bijaoui}, {Blasco}, {Bonfigli}, {Bono}, {Boudreault}, {Bressan},
  {Brown}, {Brunet}, {Bunclark}, {Buonanno}, {Butkevich}, {Carret}, {Carrion},
  {Chemin}, {Ch{\'e}reau}, {Corcione}, {Darmigny}, {de Boer}, {de Teodoro}, {de
  Zeeuw}, {Delle Luche}, {Domingues}, {Dubath}, {Fodor}, {Fr{\'e}zouls},
  {Fries}, {Fustes}, {Fyfe}, {Gallardo}, {Gallegos}, {Gardiol}, {Gebran},
  {Gomboc}, {G{\'o}mez}, {Grux}, {Gueguen}, {Heyrovsky}, {Hoar}, {Iannicola},
  {Isasi Parache}, {Janotto}, {Joliet}, {Jonckheere}, {Keil}, {Kim},
  {Klagyivik}, {Klar}, {Knude}, {Kochukhov}, {Kolka}, {Kos}, {Kutka}, {Lainey},
  {LeBouquin}, {Liu}, {Loreggia}, {Makarov}, {Marseille}, {Martayan},
  {Martinez-Rubi}, {Massart}, {Meynadier}, {Mignot}, {Munari}, {Nguyen},
  {Nordlander}, {Ocvirk}, {O'Flaherty}, {Olias Sanz}, {Ortiz}, {Osorio},
  {Oszkiewicz}, {Ouzounis}, {Palmer}, {Park}, {Pasquato}, {Peltzer}, {Peralta},
  {P{\'e}turaud}, {Pieniluoma}, {Pigozzi}, {Poels}, {Prat}, {Prod'homme},
  {Raison}, {Rebordao}, {Risquez}, {Rocca-Volmerange}, {Rosen}, {Ruiz-Fuertes},
  {Russo}, {Sembay}, {Serraller Vizcaino}, {Short}, {Siebert}, {Silva},
  {Sinachopoulos}, {Slezak}, {Soffel}, {Sosnowska}, {Strai{\v{z}}ys}, {ter
  Linden}, {Terrell}, {Theil}, {Tiede}, {Troisi}, {Tsalmantza}, {Tur},
  {Vaccari}, {Vachier}, {Valles}, {Van Hamme}, {Veltz}, {Virtanen}, {Wallut},
  {Wichmann}, {Wilkinson}, {Ziaeepour}, \& {Zschocke}}]{Gaia}
{Gaia Collaboration}, {Prusti}, T., {de Bruijne}, J.~H.~J., {et~al.} 2016,
  A\&A, 595, A1

\bibitem[{Gavel {et~al.}(2020)Gavel, Korn, Andrae, \& Fouesneau}]{iotw}
Gavel, A., Korn, A.~J., Andrae, R., \& Fouesneau, M. 2020, The chemical trace
  of {Galactic} stellar populations as seen by {Gaia},
  \url{https://www.cosmos.esa.int/web/gaia/iow_20200320}

\bibitem[{Geurts {et~al.}(2006)Geurts, Ernst, \& Wehenkel}]{ExtraTrees}
Geurts, P., Ernst, D., \& Wehenkel, L. 2006, Machine Learning, 63, 3–42

\bibitem[{Gustafsson {et~al.}(2008)Gustafsson, Edvardsson, Eriksson,
  J{\o}rgensen, Nordlund, \& Plez}]{marx}
Gustafsson, B., Edvardsson, B., Eriksson, K., {et~al.} 2008, Astronomy \&
  Astrophysics, 486, 951

\bibitem[{Helmi {et~al.}(2018)Helmi, Babusiaux, Koppelman, Massari, Veljanoski,
  \& Brown}]{gaia-enceladus}
Helmi, A., Babusiaux, C., Koppelman, H.~H., {et~al.} 2018, Nature, 563, 85

\bibitem[{Liu {et~al.}(2012)Liu, Bailer-Jones, Sordo, Vallenari, Borrachero,
  Luri, \& Sartoretti}]{gaia_spectrophotometry}
Liu, C., Bailer-Jones, C. A.~L., Sordo, R., {et~al.} 2012, Monthly Notices of
  the Royal Astronomical Society, 426, 2463–2482

\bibitem[{Majewski {et~al.}(2017)Majewski, Schiavon, Frinchaboy, Prieto,
  Barkhouser, Bizyaev, Blank, Brunner, Burton, Carrera, Chojnowski, Cunha,
  Epstein, Fitzgerald, P{\'{e}}rez, Hearty, Henderson, Holtzman, Johnson, Lam,
  Lawler, Maseman, M{\'{e}}sz{\'{a}}ros, Nelson, Nguyen, Nidever, Pinsonneault,
  Shetrone, Smee, Smith, Stolberg, Skrutskie, Walker, Wilson, Zasowski, Anders,
  Basu, Beland, Blanton, Bovy, Brownstein, Carlberg, Chaplin, Chiappini,
  Eisenstein, Elsworth, Feuillet, Fleming, Galbraith-Frew, Garc{\'{\i}}a,
  Garc{\'{\i}}a-Hern{\'{a}}ndez, Gillespie, Girardi, Gunn, Hasselquist, Hayden,
  Hekker, Ivans, Kinemuchi, Klaene, Mahadevan, Mathur, Mosser, Muna, Munn,
  Nichol, O'Connell, Parejko, Robin, Rocha-Pinto, Schultheis, Serenelli, Shane,
  Aguirre, Sobeck, Thompson, Troup, Weinberg, \& Zamora}]{APOGEE}
Majewski, S.~R., Schiavon, R.~P., Frinchaboy, P.~M., {et~al.} 2017, The
  Astronomical Journal, 154, 94

\bibitem[{Ness {et~al.}(2015)Ness, Hogg, Rix, Ho, \& Zasowski}]{cannon}
Ness, M., Hogg, D.~W., Rix, H.-W., Ho, A. Y.~Q., \& Zasowski, G. 2015, The
  Astrophysical Journal, 808, 16

\bibitem[{Pedregosa {et~al.}(2011)Pedregosa, Varoquaux, Gramfort, Michel,
  Thirion, Grisel, Blondel, Prettenhofer, Weiss, Dubourg, Vanderplas, Passos,
  Cournapeau, Brucher, Perrot, \& Duchesnay}]{scikit-learn}
Pedregosa, F., Varoquaux, G., Gramfort, A., {et~al.} 2011, Journal of Machine
  Learning Research, 12, 2825

\bibitem[{Piskunov \& Valenti(1996)}]{smeoriginal}
Piskunov, N.~E. \& Valenti, J.~A. 1996, Astronomy \& Astrophysics Supplement,
  118, 595

\bibitem[{Piskunov \& Valenti(2017)}]{smevolution}
Piskunov, N.~E. \& Valenti, J.~A. 2017, Astronomy \& Astrophysics, 597

\bibitem[{{Recio-Blanco} {et~al.}(2016){Recio-Blanco}, {de Laverny}, {Allende
  Prieto}, {Fustes}, {Manteiga}, {Arcay}, {Bijaoui}, {Dafonte}, {Ordenovic}, \&
  {Ordo{\~n}ez Blanco}}]{RVS}
{Recio-Blanco}, A., {de Laverny}, P., {Allende Prieto}, C., {et~al.} 2016,
  \aap, 585, A93

\bibitem[{scikit-learn developers(2020)}]{scikit_manual}
scikit-learn developers. 2020, User Guide,
  \url{https://scikit-learn.org/stable/user_guide.html}

\end{thebibliography}

\newpage
\begin{appendix}

\section{Performance on the GALAH sample for parameters other than $\afe$}\label{app_galah_underestimate}
In the main body of the article, we focus on estimates of $\afe$, but the models simultaneously estimate the parameters $\teff$, $\logg$, and $\feh$. To evaluate the trustworthiness of the models, we also looked at the quality of these estimates.

\begin{figure}
\centering
\includegraphics[width=8.8cm]{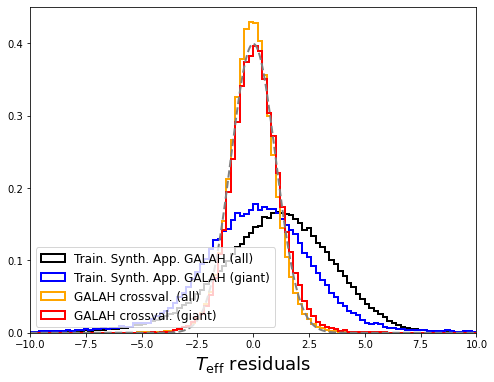}
\caption{Normalised difference, as defined in Eq.~\eqref{eq_normdiff}, between GALAH $\teff$ estimates and our estimates using the ExtraTrees algorithm. Standard normal distribution shown in grey.}
\label{ill_residuals_teff}
\end{figure}

Figure~\ref{ill_residuals_teff} shows the normalised residuals, as defined by Eq.~\ref{eq_normdiff}, for the parameter $\teff$, when applying models trained on either sample to the GALAH sample. For the cross-validation using the GALAH sample, the performance is slightly better than the theoretical maximum described in~\ref{sec_synth_crossval}. This cannot happen purely as a result of good performance on the part of the model. Rather, it shows that the errors reported by the GALAH survey are slight overestimates, at least if they are interpreted as standard deviations in normally-distributed random scatter. Limiting our results to the giants in the GALAH sample, our performance is slightly lower. With training on the synthetic sample, the scatter is considerably larger, with an offset for the entire sample but not for the subset of giants.

\begin{figure}
\centering
\includegraphics[width=8.8cm]{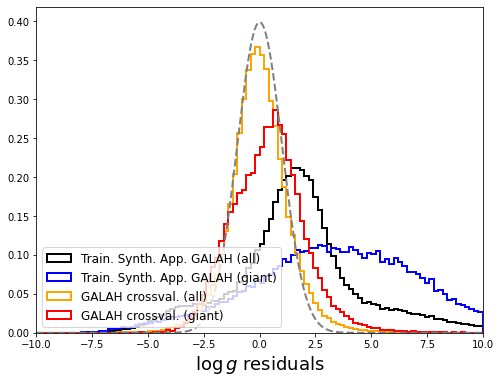}
\caption{Normalised difference, as defined in Eq.~\eqref{eq_normdiff}, between GALAH $\logg$ estimates and our estimates using the ExtraTrees algorithm. Standard normal distribution shown in grey.}
\label{ill_residuals_logg}
\end{figure}

Figure~\ref{ill_residuals_logg} shows the normalised residuals for $\logg$. The performance is close to ideal during cross-validation on the entire GALAH sample, but slightly lower for the subset of giants. With training on the synthetic sample, there is larger scatter as well as a considerable offset, which gets noticeably worse for the subset of giants.

\begin{figure}
\centering
\includegraphics[width=8.8cm]{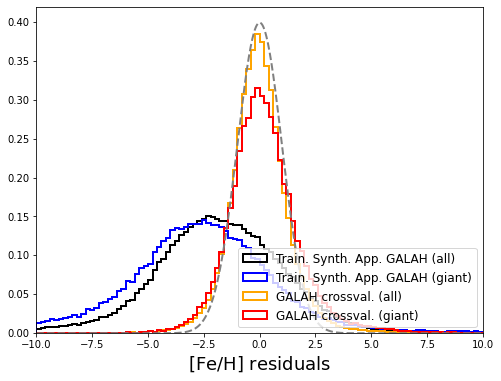}
\caption{Normalised difference, as defined in Eq.~\eqref{eq_normdiff}, between GALAH $\feh$ estimates and our estimates using the ExtraTrees algorithm. Standard normal distribution shown in grey.}
\label{ill_residuals_feh}
\end{figure}

Figure~\ref{ill_residuals_feh} shows the normalised residuals for $\feh$. The performance is again close to ideal during cross-validation on the entire GALAH sample and slightly lower for the subset of giants. With training on the synthetic sample, there is larger scatter as well as a considerable offset, but no large difference between the full sample and the subset of giants.

\section{Cross-validation on the synthetic sample with and without simulated noise}\label{app_noiseless_synth}
When performing cross-validation on the synthetic spectra in Sect.~\ref{sec_synth_crossval}, we trained the models on the synthetic spectra as they are, but then applied the model to synthetic spectra with simulated noise added. We believe this is the optimal test, since it trains the model on the highest-quality spectra we have, but does not exaggerate the performance of the model by testing it on spectra of impossibly high quality. For completeness, we also tested applying the model to spectra without simulated noise, as well as training a model on spectra with noise and then applying it to spectra with noise.

\begin{figure}
\centering
\includegraphics[width=8.8cm]{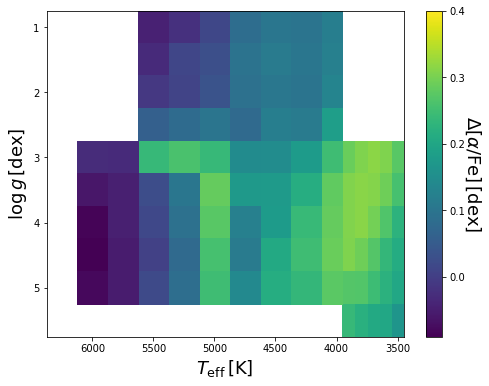}
\caption{Difference $\Delta \afe$ between the $\alpha$-rich and $\alpha$-poor stars in the synthetic sample, when performing cross-validation without adding synthetic noise. For each $\teff$-$\logg$ node, the average is taken over $\feh$ and $A_0$.}
\label{ill_synthetic_noiseless_alpha_difference}
\end{figure}

Figure~\ref{ill_synthetic_noiseless_alpha_difference} shows the $\Delta \afe$ when cross-validation is performed without adding synthetic noise to the application sample. The performance is better than that with synthetic noise in the application sample, as shown in Fig.~\ref{ill_synthetic_alpha_difference}. However, this is unlikely to reflect any performance that could be achieved with real spectra: In the dwarf sample, at $\teff = 5000\,\unit{K}$ there is a sudden rise in $\Delta \afe$, which continues along the boundary to the giant sample at $\logg = 3.0\,\unit{dex}$. This is unlikely to have any physical basis, and might indicate that the model is making use of artefacts in the synthetic spectra to estimate $\afe$. When noise is added to the spectra, this feature disappears.

\begin{figure}
\centering
\includegraphics[width=8.8cm]{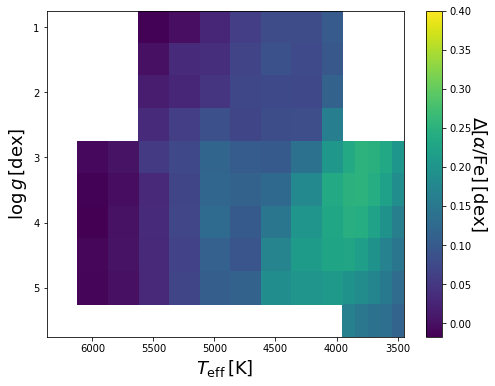}
\caption{Difference $\Delta \afe$ between the $\alpha$-rich and $\alpha$-poor stars in the synthetic sample, when performing cross-validation adding synthetic noise to both training sample and application sample. For each $\teff$-$\logg$ node, the average is taken over $\feh$ and $A_0$.}
\label{ill_synthetic_doublenoisy_alpha_difference}
\end{figure}

Figure~\ref{ill_synthetic_doublenoisy_alpha_difference} shows the $\Delta \afe$ when cross-validation is performed with the addition of synthetic noise to both the training and application sample. The performance is almost identical to that shown in Fig.~\ref{ill_synthetic_alpha_difference}, with $\Delta \afe$ shifting less than $0.03\,\unit{dex}$ in any bin.

\end{appendix}

\end{document}